\pdfoutput=1 
\documentclass[a4paper,12pt]{article}

\usepackage{amsfonts}
\usepackage[dvipsnames]{xcolor}
\usepackage{placeins}
\usepackage{graphicx}
\usepackage[perpage]{footmisc}
\usepackage[normalem]{ulem}
\usepackage{caption}
\usepackage{makecell}
\usepackage{subcaption}
\usepackage{jinstpub} 
\usepackage{multicol,tabularx,capt-of}
\usepackage{hhline}
\usepackage{booktabs}
\usepackage{multirow}
\usepackage{lineno}
\usepackage{xfrac}
\usepackage{url}
\usepackage{soul}                     

\newcommand{\direxeno}{\textsc{DireXeno}}

\title{\direxeno\ - an apparatus for measuring correlated scintillation
signatures in Liquid Xenon}

\author[a,b,1]{R. Itay,\note{Corresponding author}}
\author[a]{P. Z. Szabo,}
\author[a]{G. Koltman,}
\author[a,c]{M.M. Devi,}
\author[a]{M. Shutman,}
\author[a]{H. Landsman,}
\author[a,d]{R. Budnik}

\affiliation[a]{Department of Particle Physics and Astrophysics Weizmann Institute of Science,\\Herzel St. Rehovot, 7610001, Israel}
\affiliation[b]{SLAC National Accelerator Laboratory,\\ Sand Hill Rd. Menlo Park, Ca, 94025, USA}
\affiliation[c]{Department of Physics, Tezpur University,\\Tezpur University Rd., Assam, 784028, India}
\affiliation[d]{Department of Physics and Astronomy, Simons Centnter for Geometry and Physics and C. M. Yang Institute for theoretical physics. Stony Brook NY, 11794, USA}
\emailAdd{ranitay@stanford.edu}
\abstract{We present a detector apparatus, \direxeno\ (DIREctional XENOn), designed to measure the spatial and temporal properties of scintillation in liquid xenon to very high accuracy. The properties of scintillation are of primary importance for dark matter and neutrinoless double beta decay experiments; however the complicated microphysics involved limits theoretical predictions. We will explore the possibility that scintillation emission exhibits spatial correlations such as super-radiance which manifests in temporal and spatial structure, depending on the interaction type. Such properties of scintillation light may open a new window for background rejection as well as directionality measurements.  We present the apparatus' technical design and the concepts driving it. We demonstrate that for an energy deposition of $\sim2.5$\,keV ($\sim7.5$\,keV) electron (nuclear) recoil the detector is sensitive to an anisotropy fraction of as little as $\sim$ 20\% of the total photons emitted over a solid angle of $\sim0.85$ steradian or less.
We show results from commissioning runs in which the detector operated with 17 PMTs for over 44 days in stable conditions. The time resolution for individual photons in different PMTs was measured to be $\lesssim1.4$\,ns full-width at half-maximum}


\begin{document}
\maketitle

% ========================================================================
% introduction
% ========================================================================
\section{Introduction}
\label{sec:intro}

Noble-Liquid detectors have gained popularity over the past decades.
Specifically, liquid xenon (LXe) is widely used as the target material in dark matter (DM) direct detection and neutrino-less double beta-decay ($0\nu\beta\beta$) experiments. The scintillation light and the ionization electrons produced in an interaction are used to reconstruct the recoil energy, the interaction location, and to discriminate between nuclear recoils (NR) and electron recoils (ER) \cite{xenon1t,Akerib:2016vxi, PandaXresults, Abe2018_xmass,Auger_2012}. Some experiments use a dual phase time projection chamber (TPC) based detector that measures both the xenon ionization and scintillation signals, and may thus provides better location reconstruction and ER/NR discrimination.

The current primary backgrounds in LXe TPCs are ERs from $\gamma$ and $\beta$ emission by residual radioactivity in the detector and its environment. This includes both impurities in the LXe and radioactive isotopes in the material that constructs the detector. The next generation of DM experiments is planned to be sensitive enough to reach the \emph{neutrino floor} \cite{Billard:2013qya,Aalbers:2016jon}, where neutrinos from various sources produce a substantial rate of irreducible NR background with respect to the expected signal rate.

When a particle interacts with the LXe it produces a recoil electron or nucleus. The recoiling particle forms a set of excited xenon atoms (Xe$^*$, \emph{exciton}) and electron-ion pairs (Xe$^{+} +$ e$^{-}$). Excitons form excited dimer states (Xe$^*_2$, \emph{excimer}) through interaction with neighbouring xenon atoms, and these excimers emit VUV light ($h\nu$) when they decay from the lowest excited molecular state to the dissociative ground state,
\begin{equation} \label{eq:XeSci1}
    \text{Xe}^*+\text{Xe} \rightarrow \text{Xe}^*_2 \rightarrow 2\text{Xe} + h \nu.
\end{equation}
Some of the ionized atoms and free electrons recombine to form excitons,
\begin{equation} \label{eq:XeSci2}
\begin{split}
  &\text{Xe}^{+} + \text{Xe} \rightarrow \text{Xe}^{+}_2 \\
  &\text{Xe}^{+}_2 + e^{-}  \rightarrow \text{Xe}^{**} \\
  &\text{Xe}^{**} \rightarrow \text{Xe}^{*} + \text{heat} \\
  &\text{Xe}^{*} + \text{Xe} \rightarrow \text{Xe}^{*}_2 \rightarrow 2\text{Xe} + h\nu,
\end{split}
\end{equation}
and the decay of the resulting excimers also contributes to the VUV emission \cite{Doke:1988rq}.

The scintillation spectrum has a Gaussian shape centered at a wavelength of 175~nm with a 10~nm full-width at half-maximum (FWHM) \cite{LXeWL}. The excimers have two decay components with different temporal structure, one from the singlet  and the other from the triplet state. In the absence of an external electric field, ER induced excimer production by recombination (eq.~\ref{eq:XeSci2}) dominates over the singlet and triplet emission with a decay time of $\sim$ 45~ns. For NRs the recombination process is very fast, less then 1~ns, and both the singlet and triplet components can be observed. The singlet state decay time is $\sim$(2 - 4)\,ns, and the triplet state decay time is $\sim$(24 -27)\,ns \cite{LXEdecay}.

The phenomenon of spontaneous phase-locking of atomic dipoles in an excited sample, known as super-radiance (SR)~\cite{SR}, opens new prospects for background rejection in liquid xenon detectors. SR occurs when the excited sample size is smaller than the radiation wavelength, such that the photon emission cannot be assigned to a specific emitter in the sample. In ordinary radiance from an excited sample of atoms the emission pattern is isotropic, the decay time is exponential with a characteristic time $\tau_R$, and the intensity is proportional to the number of emitters ($\text{I}_\text{R}\sim$ N); in SR the excited sample radiates coherently ($\text{I}_{\text{SR}}\sim$ N$^2$) much faster ($\tau_{\text{SR}}\sim\tau_\text{R}/\text{N}$) and the radiation is enhanced in the direction of the longest dimension of the sample. It has been shown by simulations that the typical size of NR events is $\mathcal{O}(100 \mathrm{\,nm})$~\cite{NRSize} which is smaller than the xenon radiation wavelength ($\sim$175~nm). The expected size of the excimer cloud for ER is $\sim$700\,nm~\cite{Aprile:2006kx}; moreover, in the absence of external electric field the ER emission time structure is dominated by recombination which further reduces the probability for SR and thus SR may be an identifier of NR.

The directionality of the SR emission holds a potential for the rejection of the irreducible neutrino background, which poses a challenge for future experiments. The directions of the NRs induced by solar neutrinos are peaked in the direction away from the Sun, while the diffuse Supernovae and atmospheric neutrinos NRs are isotropic~\cite{NeutrinoBG1, NeutrinoBG2}. The directions of DM induced NRs are peaked opposite to the direction of the solar system velocity vector around the center of the Milky Way and result in a forward-backward asymmetry in the recoil rates in the Galactic reference frame~\cite{DMDirection}. No known background can mimic this signal, and the directional signal is widely held to be the cleanest expected signature of Galactic DM. 

Early studies of macroscopic ionization using high energy-density electron beams showed that scintillation from LXe produces a coherent amplification of light~\cite{Basov}. However, SR emission from a sample excited by a single particle is yet to be explored. In this paper we present the set-up of an experiment called \direxeno\ (DIREctional XENOn), specifically designed to measure the spatial and temporal distribution of LXe scintillation light from a localized event.

% ========================================================================
\section{Experiment Setup}
% ========================================================================
\label{sec:exp_setup}

\begin{figure}[htb]
\centerline{\includegraphics[width=1\linewidth]{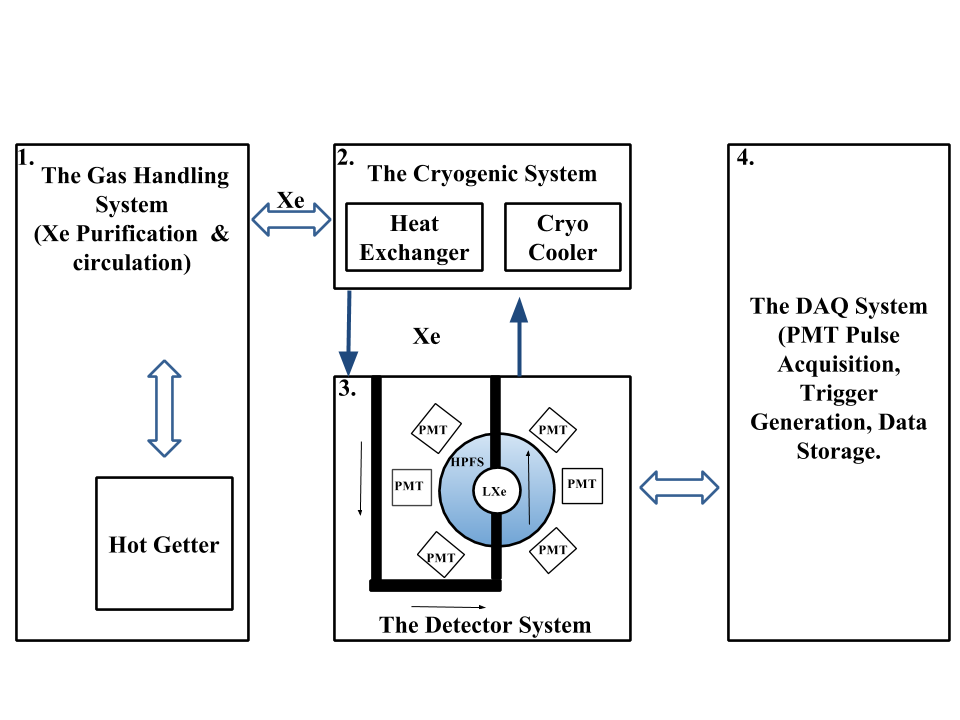}}
\caption{A schematic view of \direxeno. The slow control parts are distributed around all subsystems}
\label{fig:detSch}
\end{figure}

In the heart of \direxeno\ lies a small spherical cavity inside a thick sphere made of high purity fused silica (HPFS). LXe is circulated through the cavity and serves as a uniform excitation-target. The sphere is surrounded by 20 photomultiplier tubes (PMTs) facing its center, allowing spatial and temporal measurements of individual photons. The geometry of the detector approximates a point source of scintillation photons, so detailed vertex reconstruction inside the LXe is unnecessary. A schematic view of this system is shown in figure~\ref{fig:detSch}.

The system is designed to achieve $\sim1$\,ns synchronization between PMTs, and $\sim 0.5$\,sr angular resolution. Since the nature and magnitude of SR in LXe are yet unknown, the guiding principles of the design were flexibility and versatility. Any part of the system can be redesigned or upgraded to fulfill future experimental requirements without changing the rest of the system. The modular design also allows fast and easy recovery in case components malfunction.

\begin{figure}[htb]
\centerline{\includegraphics[width=1.\linewidth]{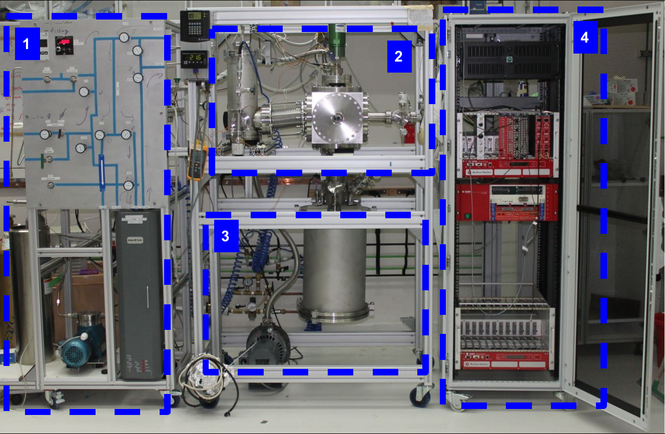}}
\caption{\direxeno\ mounted on the three racks: $(1)$ the gas handling system; $(2)$ the cryogenic system, including the heat-exchanger; $(3)$ the detector assembly; and $(4)$ the data acquisition system. The slow-control is distributed around all three racks.
\label{fig:fulldet}}
\end{figure}

\direxeno\ is made of five main building blocks: $(i)$ the \emph{gas handling system} provides a xenon filling and recovery port, circulates the xenon and purifies it; $(ii)$ the \emph{cryogenic system} liquefies the xenon and delivers it to the detector; $(iii)$ the \emph{detector assembly} holds the HPFS sphere and the PMTs around it; $(iv)$ the \emph{data acquisition system} (DAQ) supplies high voltage to the PMTs and handles triggering and digitization of data; and $(v)$ the \emph{slow control system} (SC) monitors the condition of the experiment using various sensors and gauges. The entire assembly is held on three separate racks as shown in figure~\ref{fig:fulldet}.

% ========================================================================
\subsection{Gas Handling}
% ========================================================================
\label{subsec:gas_sys}

In LXe detectors which measure the ionization electrons the desired level of impurity concentration is 1\,ppb O$_2$ equivalent or better~\cite{xenon1t,Akerib:2012ys}. In order for the LXe to reach this level of purity, continuous purification is required. In \direxeno\ only the prompt scintillation is measured, so such a high level of purification is not necessary. Nonetheless, this capability along with filling, circulation and recovery of xenon is provided by the gas handling system.

\begin{figure}[htb]
\centerline{\includegraphics[width=1\linewidth]{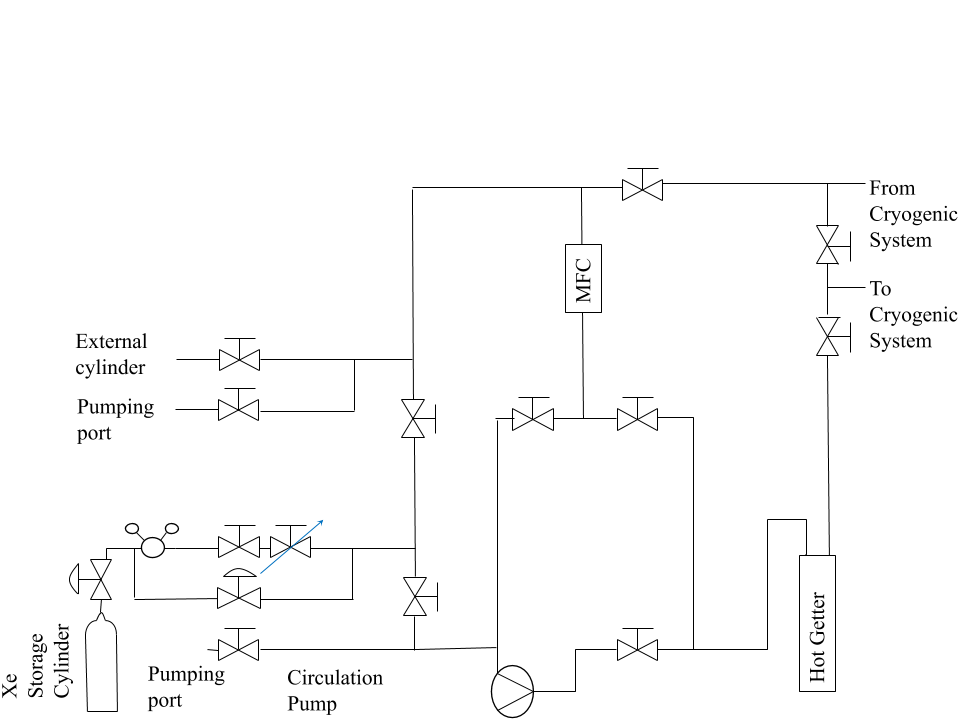}}
\caption{Gas handling system piping and instrumentation diagram. Arched taps indicate high-pressure valves and arrows indicates needle valves
\label{fig:gasSchematic}}
\end{figure}

During purification, a circulation pump\footnote{KNF N143 SN.12E diaphragm-gas sampling pump.} extracts LXe from the detector assembly through a heat exchanger\footnote{GEA GBS100M 24 plate heat-exchanger.}, where it is heated and vaporized. The xenon then passes through a Mass Flow Controller\footnote{MKS mass flow controller 1179A00614CR1BM TBD} (MFC) that can regulate the heat flux into the system and pumped into a hot getter\footnote{SAES MonoTorr PS4-MT15-R-2 heated getter purifier.} that removes residual carbon, nitrogen, water and other oxygen based impurities to the level of $<1$\,ppb (according to product specifications). Once purified, the gaseous xenon (GXe) is delivered back to the cryogenic system via the heat exchanger, where it is liquefied and directed into the detector assembly. A schematic of this system is shown in figure~\ref{fig:gasSchematic}.

% ========================================================================
\subsection{Cryogenics}
% ========================================================================
\label{subsec:cryo_sys}

Remote cooling is commonly used in LXe experiments, assisting in reduction in background radiation and acoustic noise from the cooler to the detector, and also providing design flexibility. The cryogenic system is connected to the gas handling system on one side and to the detector assembly on the other, and built such that replacing the cryo-cooler (e.g., to a pulse tube refrigerator) requires just an adapter at the top flange.

The cryogenic system is divided into an outer vessel (OV) which holds the insulation vacuum, and an inner vessel (IV) which holds the xenon. In addition to the vacuum jacket that minimizes diffusive and convective heat leaks, the IV is fully covered with a multi-layer aluminized mylar to reduce radiative heating.  

\begin{figure}[htb]
\centerline{\includegraphics[width=1\linewidth]{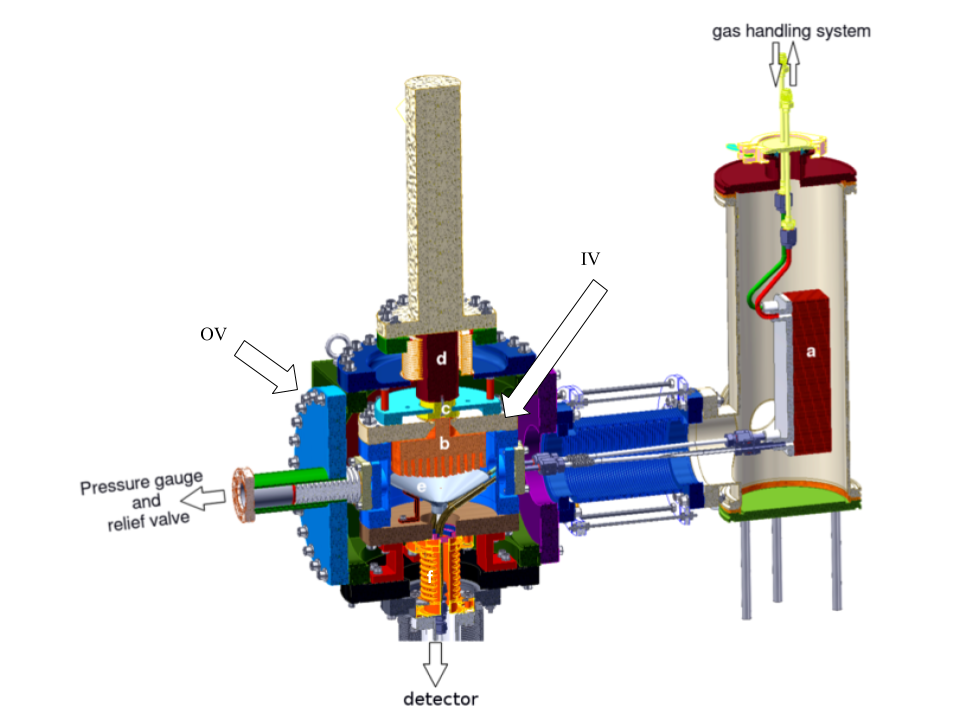}}
\caption{CAD view of the cryogenic system. The interfaces to the detector part, the gas handling system and the relief valves are indicated by the arrows as well as the outer vessel (OV) and inner vessel (IV). The parts are: a) heat exchanger, b) cold finger, c) copper adapter thermally connecting the cold finger to the cryo-cooler, d) cryo-cooler, e) funnel, f) bellows to detector part.} 
\label{fig:cryo}
\end{figure}

The OV is made of a 10"  ConFlat (CF) cube, with ports on all six faces, interfacing the gas handling system and the detector assembly, and bearing various service ports (e.g., feed-throughs, pumping ports, view ports). The OV is connected to the detector assembly via a 6" CF flexible bellows, providing a shared vacuum jacket. 

The IV is made of an 8" CF nipple with a 6" length. A 120\,~mm diameter oxygen-free, high thermal-conductivity (OFHC) copper cold finger is welded to its top flange with a design similar to the one in~\cite{xenon100}. The inner part of the cold finger, which is in contact with the xenon, has long fins to maximize its surface area and hence its heat-transport flux. The upper part of the cold finger is in thermal contact with the bottom part of a cryo-cooler\footnote{QDrive 2S132K-22690-A.} via a copper adapter. A cartridge heater inside the copper adapter can be used for emergency heating in case xenon freezes on the cold finger. The cryo-cooler itself is mounted on a $4~\sfrac{1}{2}$" feed-through on the OV top flange. It is controlled by a drive electronics unit\footnote{Allen-Bradley 20BB9P6A3-AYNBNC0 with an integrated OMEGA temperature control unit.} that allows setting the target cooling temperature and maintaining it stable to $\lesssim 0.2^{\circ}$C by a PID control loop. This system has the advantage that, unlike many other LXe experiments which use a heater to regulate the temperature, this controller regulates the output power of the cryo-cooler (up to a maximum of about 70\,W cooling power) to maintain the desired temperature in varying conditions.

A custom-made 0.6\,mm thick stainless steel (SS) funnel is installed on the inner side of the IV lower flange, collecting LXe drops from the cold finger and delivering them to the detector. The IV lower flange is attached to the detector assembly via a $3~\sfrac{3}{8}$~" flexible bellows, creating a continuous volume that allows the GXe evaporated in the detector to reach the cold finger and recondense. The bellows also hosts three pipes: the suction and return pipes from the gas handling system, and the pipe coming from the funnel.

The purified LXe from the gas handling system and the less pure LXe (from the cold finger) run in separate pipes, and can be delivered to different parts of the detector. The CAD view of the cryogenic system design is shown in figure~\ref{fig:cryo}. 

% ========================================================================
\subsection{The Detector}
% ========================================================================
\label{subsec:det_sys}
 
\begin{figure}[htb]
\centerline{\includegraphics[width=1\linewidth]{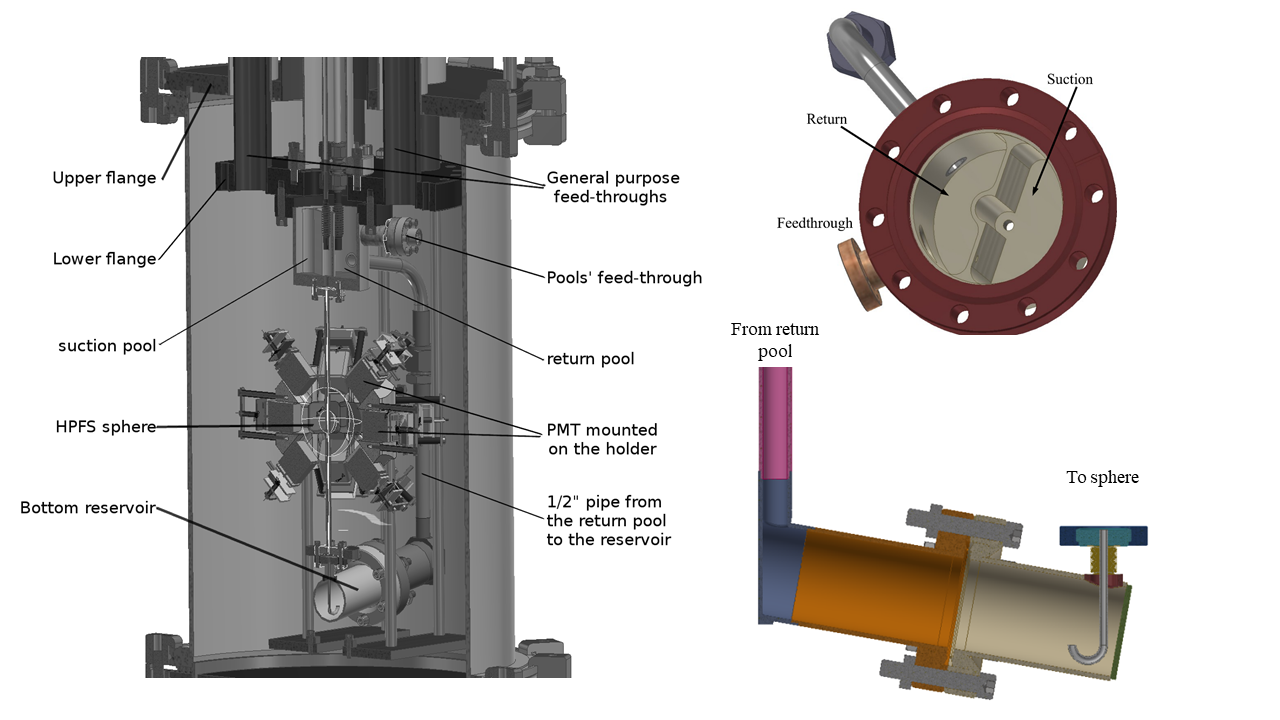}}

\caption{(Left) A CAD view of the detector assembly. (Right top) Top view of the upper pools (Right bottom) The bottom LXe reservoir }
\label{fig:detector}
\end{figure}

The detector assembly is a vacuum chamber with an inner assembly consisting of the HPFS sphere, the PMT sensors observing it and their accessories. This chamber is placed below the cryogenic system and can be seen in figure~\ref{fig:detector}. 

The interface unit to the cryogenic system consists of two flanges welded together via seven tubes, which serve as general purpose service ports, four ports with a $2~\sfrac{3}{4}$" CF flange, and three with a $1~\sfrac{1}{3}$" CF flange (mini-CF). The upper flange (NW320 ISO-LF) shares the cryogenic system's OV insulation vacuum, while the lower one (10" CF) is part of the IV and  can hold a full size LXe vessel in a future detector upgrade. The lower flange of the interface unit (see  figure~\ref{fig:detector}) is also adapted to fit a smaller 4~\sfrac{5}{8}" CF flange, which is used in the current detector configuration to connect the pools assembly (described below), see figure~\ref{fig:detector}. 

The vacuum chamber is made of an NW320 ISO-LF nipple closed with a blank flange from below, and connected to the upper interface flange from above. The length of the nipple is designed to be 50~cm. The height of the setup is less than 190~\,cm, allowing transporting the experiment for irradiation. 

The $4~\sfrac{5}{8}$" CF adaptation of the interface unit's lower flange  holds a split vessel with two pools (see figure~\ref{fig:detector}) that serves as a LXe reservoir. One pool (suction) is connected to the top port of the HPFS sphere, and the other (return) to its bottom. The circulation path is such that LXe coming from the funnel, or directly from the gas handling system, drips into the return pool, flows through a $\sfrac{1}{2}$" pipe down to a bottom reservoir, then up through the sphere and into the suction pool. From the suction pool it is pumped back to the gas handling system. This creates a buffer that maintains the sphere constantly filled with LXe, and ensures that the flow and the convection in the LXe are in the same direction. Each pool is equipped with two temperature sensors\footnote{Lakeshore PT111.} for liquid level monitoring (see also section~\ref{subsec:sc}). The wires of the sensors are connected through a mini-CF feed-through to the OV (see figure~\ref{fig:detector}). 

The bottom LXe reservoir (see figure~\ref{fig:detector}) is connected below the sphere and serves as a xenon phase-separator and a thermal buffer. Its inclined design prevents bubbles that form inside it from mixing into the up-flowing fluid. These bubbles accumulate in the top part of the reservoir and eventually float back to the return pool through the $\sfrac{1}{2}$" input pipe (see figure~\ref{fig:detector}) rather than through the sphere. The inner pipe feeding the sphere from the liquid phase inside the reservoir is curved upwards, so bubbles that form below its opening will not reach the sphere. This way only a single-phase fluid flows through the sphere and scintillation light is not refracted by liquid-gas boundary layers inside it.

\begin{figure}[htb]
\centering
\begin{subfigure}[c]{0.4\textheight}
\includegraphics[width=\textwidth]{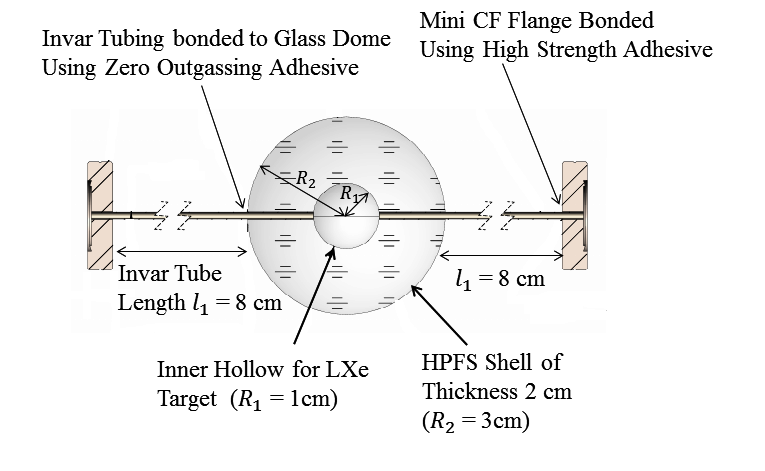}
\end{subfigure}
\begin{subfigure}[c]{0.25\textheight}
\includegraphics[width=\textwidth]{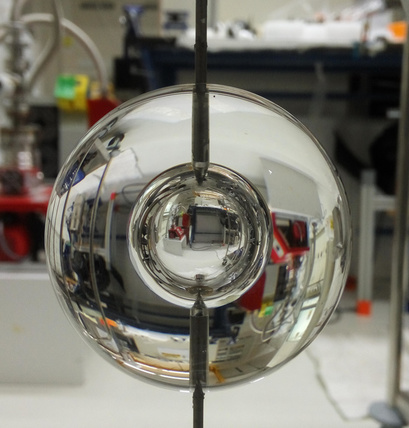}
\end{subfigure}
\caption{(Left) The technical design of the HPFS shell with Invar tubing and mini CF flanges. 
(Right) The industrially manufactured HPFS shell.} 
\label{fig:sphere}
\end{figure}

The sphere is a custom-designed hollow shell made of Corning HPFS 8655 with high transmittance to VUV. Invar tubes with SS mini-CF flanges are connected to the sphere on both sides, serving as input~/~output ports (see figure~\ref{fig:sphere}). The tubes are glued into the two holes in the sphere using a low out-gassing, cryogenic epoxy resin\footnote{Master Bond EP29LPSP}. The bottom flange of the sphere assembly is connected to the lower thermal bath via a small flexible bellows in order to prevent torque on the glued ports and to absorb vibrations. The optical properties of the sphere are further discussed in section~\ref{sec:sphere_prop}. 

Photons emitted from the LXe in the sphere are detected by 20 PMTs\footnote{Hamamatsu R8520-406 1" PMT, active area 20.5\,mm $\times$ 20.5\,mm}. The PMTs were tested by the manufacturer in room temperature to have a quantum efficiency (QE)~$\geq~30\%$, which implies a QE $\geq 33\%$ at 170\,K~\cite{Aprile:2012dy} at a wavelength of 178 nm. A positive voltage divider\footnote{Hamamatsu E13416 MOD.} is used to provide high voltage to the PMTs, in the range of 700--900 V.

\begin{figure}[htb]
\centering
\begin{subfigure}[c]{0.35\textheight}
\includegraphics[width=\textwidth]{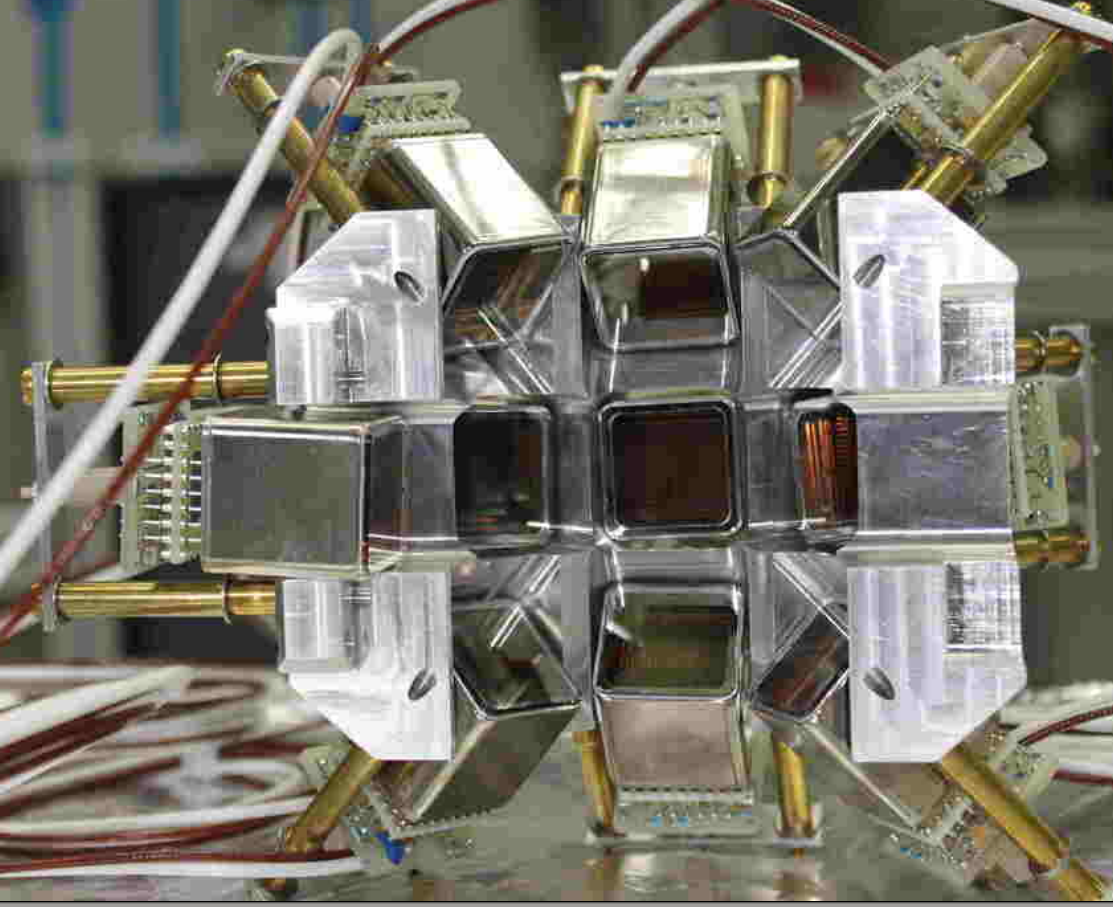}
\end{subfigure}
\begin{subfigure}[c]{0.3\textheight}
\includegraphics[width=\textwidth]{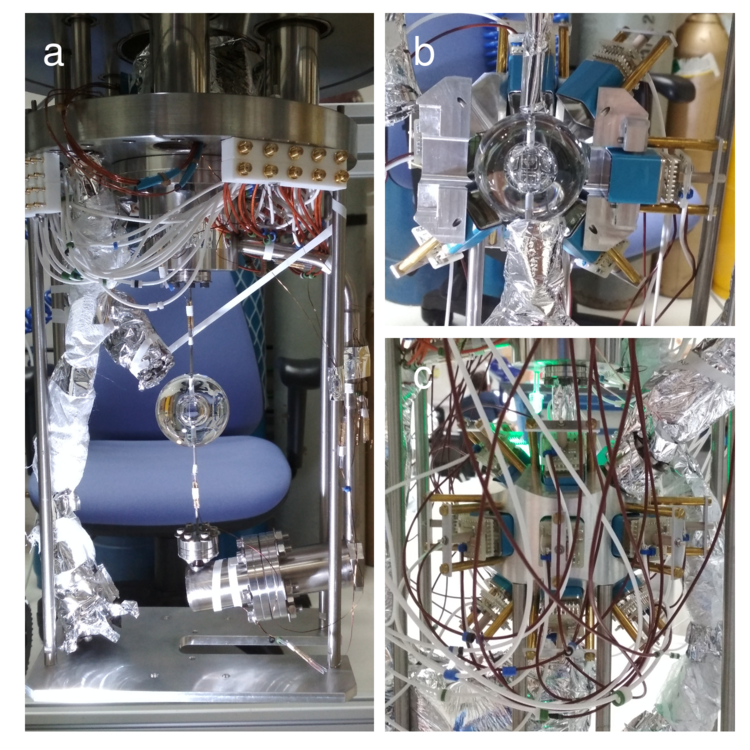}
\end{subfigure}
\caption{(Left) One hemisphere of the PMT holder. Two identical such hemispheres are used to hold the PMTS around the sphere. (Right) photos of the detector assembly without the PMT holder (a), with partial installation of the PMTs (b), and complete with PMTs installed (c) } 
\label{fig:pools_holder}
\end{figure}

The PMTs are held with a custom-designed aluminum holder. The holder is made of two hemispheres hosting the PMTs in 3 rows, all of them pointing to the center of the sphere. The PMTs are attached to the holder by their voltage--divider bases using M2 PEEK screws. In figure~\ref{fig:pools_holder} the PMT assembly (holder and PMTs) is presented as well as the bottom reservoir, pools part, and the sphere.

In some of the initial tests a USB snake  camera was inserted into the system for monitoring the xenon in the sphere. The camera was inserted through a mini-CF port connected to a \sfrac{1}{4}" flexible tube with a viewport on its other side. Thus the camera was never in vacuum or exposed to cold temperatures. 

Finally, an optical fiber with a diameter of 1000~$\mu$m and numerical aperture of 0.22, leads 900 ps long (FWHM) light pulses with a wavelength of 365 nm from an outer pulsed-LED light source\footnote{Edinburgh Instruments EPLED-365} to the vicinity of the sphere. Along with the light pulse itself, the LED driver-circuitry also sends a synchronized trigger signal to the data acquisition system. These two outputs are used for gain and timing calibrations (see sections~\ref{subsec:gain} and~\ref{subsec:temp_res}).

% ========================================================================
\subsection{Data Acquisition}
% ========================================================================
\label{subsec:daq}

\begin{figure}[htb]
   \centering
   \includegraphics[width=0.85\textwidth]{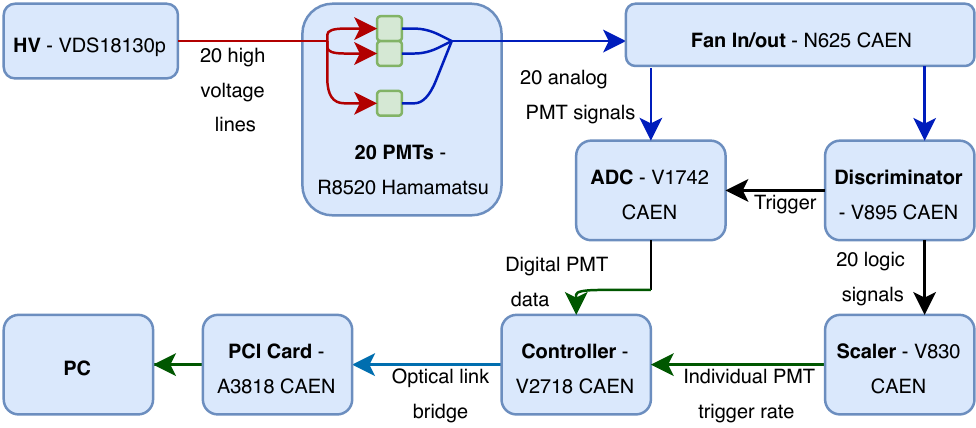}

   \caption{The schematic of the data acquisition system of \textsc{DireXeno}. The signal coming from 20 PMTs and the subsequent electronic channels to record the events once triggered. 
}
   \label{Fig:DAQscheme}
\end{figure}

The DAQ system uses both NIM and VME electronic modules. The data are read out through a PCIe card\footnote{CAEN A3818 PCIe card.}, which is connected by an optical link to a VME controller\footnote{CAEN V2718 VME controller.}. A schematic layout of the DAQ system is shown in figure~{\ref{Fig:DAQscheme}}.

The PMTs are ramped up to their individual working voltage using a VME high voltage distributor module\footnote{Iseg VDS18130p 24 independent channels positive polarity voltage distributor.}. The raw pulses from the PMTs are split by a NIM fan-in fan-out (FIFO) module\footnote{CAEN N625/N454/N405 FIFO/Logic units.} before they are processed by the NIM logic and digitized. One of the outputs is sent to a waveform digitizer ADC\footnote{CAEN ADC V1742 switched capacitor digitizer.}, and the other is fed to a discriminator\footnote{CAEN V895 16 channel leading edge discriminator.} which produces a trigger signal to the ADC when a predefined number of PMTs cross a threshold in an adjustable time window. This discriminator is also connected to a  scaler module\footnote{CAEN V830 16 channel scaler.} to monitor the individual PMT trigger rate (see section~\ref{subsec:sc}).

The ADC is a 32 channel, 12 bit waveform digitizer based on switched capacitor arrays. It has a maximum sampling rate of 5GS/s, a bandwidth of 600 MHz, an input range of 1V peak-to-peak with an adjustable offset, and a maximum event record length of 1024 samples. After the waveforms are digitized they are stored as binary files on a local drive and backed up on a network drive.

% ========================================================================
\subsection{Slow Control}
% ========================================================================
\label{subsec:sc}

A variety of sensors are used to collect information from the experiment's subsystems. This information is monitored to ensure the stability and well-being of the system as well as to understand the xenon flow conditions. A time-series server  was built specifically for handling time-stamped events  and measurements, based on influxdb~\cite{influxDB}. Grafana~\cite{Grafana}, an open-source software, is used for monitoring and visualization. The monitored data is streamed to the database using the influxdb API, which is integrated using python control scripts.

Ten temperature sensors continuously monitor the temperature of different components in the experiment: two\footnote{LakeShore PT-111} in the copper adapter above the cold finger, four\footnote{Cryocon GP-100.} in the upper pools for liquid level monitoring and another four\footnote{Cryocon XP-100.} on the IV tubes of the detector. These sensors are connected to a reader\footnote{Cryocon 18i cryogenic temperature monitor.} that provides both display and data connectivity to the SC computer. The vacuum in the inner and outer vessels is measured by two vacuum gauges\footnote{Pfeiffer Vacuum PKR 251, FPM Sealed, DN 40 ISO-KF.}. The gas pressure in the IV is monitored by a manometer\footnote{MKS 722B Baratron transducer.} connected to a readout and control unit\footnote{MKS PDR2000.}. If the IV pressure exceeds the HPFS sphere's specifications (maximal inner pressure of 3 bar gauge) a pressure relief system reduces the inner pressure by releasing xenon to the room, until pressure reaches the desired level.

Four thermistor sensors monitor the inlet and outlet ports of the cooling-water, the cryo-cooler's compressor tank and the ambient air temperature. They are read using an arduino board with an accuracy of $\sim 1^{\circ}$C. The total amount of xenon in the system is monitored using the MFC. The voltage, the current and the trigger-rate of the PMTs are monitored and streamed to the SC database as well.

A $\sfrac{1}{4}$" vacuum-tight bellows enters the OV through a mini-CF service port, with an optical view-port at its end. It holds a commercial USB endoscope camera, placed instead of one of the upper-row PMTs such that it points at the sphere. This camera allows a continuous on-the-fly visual inspection of the xenon phases inside the sphere. A simple and fast image processing algorithm constantly monitors the difference between two consecutive camera snapshots which is then streamed to the SC database.

Finally, the database is continuously checked for missing information, which may occur in case of gauges failure, connection errors or power failures. It is also checked for abnormal working conditions, such as excessive pressure and temperatures. An alarm message is sent in each of these cases, via emails and SMS.

% ========================================================================
\section{Optical Properties of the Sphere }
% ========================================================================
\label{sec:sphere_prop}

The central component of the experiment is the HPFS sphere, which holds the LXe target, located in the center of the detector assembly. In order to allow the measurement of the original direction of photons emitted by the LXe, it is important to reduce the deflection and absorption of photons on their path from the LXe to the PMTs.   

\begin{figure}
    \centering
    \includegraphics[width=1\linewidth]{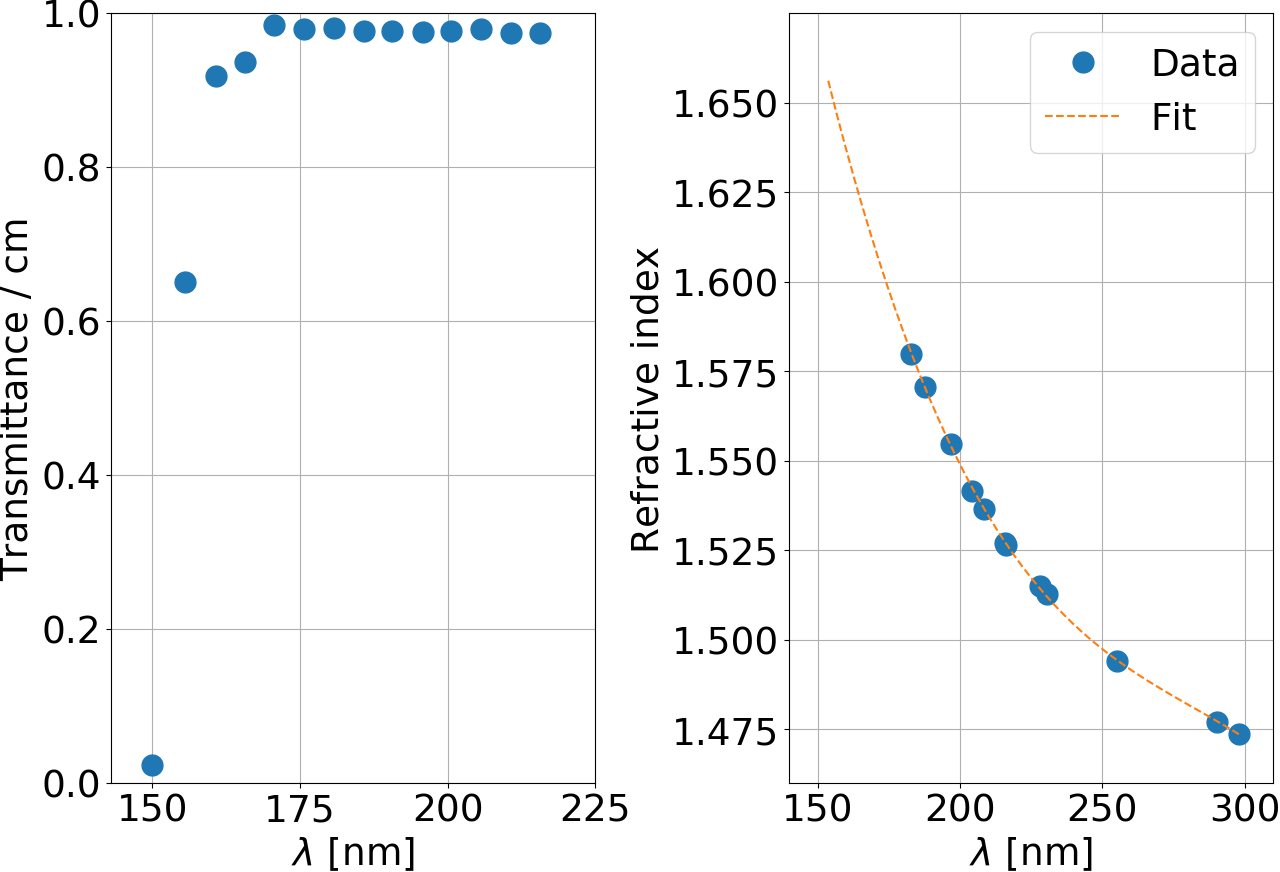}
    \caption{Corning HPFS-8655 characteristics. The internal transmittance $T_{i}$ measured on an HPFS sample (left); and the refractive index, as measured by the manufacturer, with a 3rd-order polynomial fit (right) The relevant wavelength for LXe is $\sim178$\,nm. 
    }
    \label{fig:hpfsRIcalibration}
\end{figure}

The HPFS transparency to VUV photons is a crucial parameter for setting the dimensions of the sphere (inner and outer radii). Therefore, the transmittance of an HPFS sample was measured using a VUV monochromator\footnote{McPherson 234/302VM.}. A deuterium VUV light-source\footnote{McPherson 632.} was set facing a vacuum chamber with a PMT inside, and an HPFS sample was placed in front of the PMT window. The light intensity was measured by the PMT with and without the HPFS sample, and the HPFS transmittance was calculated from the ratio of the two measurements taking into account the Fresnel reflections on both sides of the sample. The transmittance as a function of wavelength is shown in figure~\ref{fig:hpfsRIcalibration}~(left), and is $\sim98.7\%$ per cm at 178~\,nm. 

The HPFS refractive index at 178\,nm is 1.6 which matches the refractive index of LXe at 178\,nm, 1.56 -1.69 ~\cite{Solovov:2003ax,Barkov:1996hc,doi:10.1002/pssb.2221430239}. The refractive index at various wavelengths was provided by the manufacturer and is presented in figure~\ref{fig:hpfsRIcalibration}~(right). 

The sphere's inner ($r_\text{in}$) and outer ($r_\text{out}$) radii were designed to minimize the deflection of photons (higher $r_\text{out}/r_\text{in}$) and for minimal losses (lower $r_\text{out}-r_\text{in}$). GEANT4-based simulations assisted in the choice of $r_\text{in} = 1$~cm and $r_\text{out} = 3$~cm. The relevant geometrical and optical parameters, which are used in the simulation, are listed in table~\ref{tab:OptPar}.

% ========================================================================
\section{Detector Sensitivity}
% ========================================================================
\label{sec:det_sens}

In this section we estimate the sensitivity to the emission anisotropy level. The sensitivity of the detector is defined by its ability to identify an anisotropic component in the scintillation photons emission pattern on top of the isotropic one. The photon emission pattern is modeled using a single gaussian beam on top of an isotropic component :

\begin{equation}
\label{eq:aniso_pattern}
\mathcal{F}(\theta,\phi) = (1-R_\text{aniso}) \cdot f_\text{iso} + R_\text{aniso}\cdot f_\text{G}(\mu(\theta_0,\phi_0),\sigma),
\end{equation}

where $f_\text{iso}$ is the Probability Density Function (PDF) of an isotropic emission, $f_\text{G}$ is a PDF of a Gaussian distribution with a mean $\mu$ and a half-width at half-maximum (HWHM) of $\sigma$, $R_\text{aniso}$ is the anisotropic emission fraction. 
The beam direction $(\theta_0, \phi_0)$ is random and each pattern is characterized by $R_\text{aniso}$ and $\sigma$. 
These types of variations are selected as light from SR is expected to be focused along the long symmetry axes of the excimers cloud.

In each simulation event, photons are generated at the center of the LXe sphere, and  propagated through the detector until they hit a PMT or escape the system.  The number of generated photons for each event is a random number drawn from a Poisson distribution with a mean of 60, which implies an energy deposition of $\sim2.5\,\mathrm{keV}$  ($\sim7.5\,\mathrm{keV}$) for ER (NR).
The photon initial direction is randomly drawn from eq.~\ref{eq:aniso_pattern}.  
The QE at 170\,K is taken to be 33\% (see section~\ref{subsec:det_sys}). The first dynode collection efficiency (CE) is taken to be $81\%$~\cite{LiorPMT}. 
The R8520 PMTs have a 20\% probability for double photo electron (DPE) emission on conversion of 178 nm photons~\cite{Faham:2015kqa}, which is included in the simulation. Hence a photon reaching the PMT has a 27\% probability to be detected, 53\% probability to be absorbed and 20\% probability to be specularly reflected. Taking into account these probabilities as well as the geometrical properties of the system (see section~\ref{subsec:det_sys}); on average only 7 photons are detected out of the initial 60 photons produced. The DPE effect can artificially mimic a jet--like component in a purely isotropic sample, especially in events with a few detected photons, which is the case in our study.

For each event, the angles between all possible pairs of PMTs are calculated and weighted according to the number of observed hits. This correlation pattern is then summed over many events, thus disregarding the jet's main direction ($\theta_0$,~$\phi_0$) which can vary between events. In order to quantify the anisotropy of the emission, the angle correlation distribution is compared to that of an isotropic pattern using a $\chi^2$ test. It was verified that this null hypothesis follows a $\chi^2$ distribution.

\begin{table}[htb]
  \centering
     \caption{The parameters used in simulation}
  \begin{tabular}{|l c||l c|}
  \hline
  Parameter & Value & Parameter & Value \\
  \hline
  \hline
  LXe absorption length & 100 cm & HPFS shell inner radius & 1cm \\
  LXe scattering length & 35 cm & HPFS shell thickness & 2 cm\\
  LXe refractive index & 1.61  & PMT efficiency &  27\% \\
  LXe Scintillation wavelength & 178 nm& PMT distance from center & 40 mm\\
  HPFS absorption length & 100 cm  & Number of PMT & 20 \\
  HPFS refractive index & 1.57 & PMT active area & 20.5mm $\times$ 20.5mm \\
  HPFS scattering length & $\infty$ & Invar tube diameter & 2 mm\\
  DPE fraction & 0.2 & & \\ 
  \hline
 \end{tabular}
  \label{tab:OptPar}
\end{table}

\begin{figure}[htb]
\centerline{\includegraphics[width=1\linewidth]{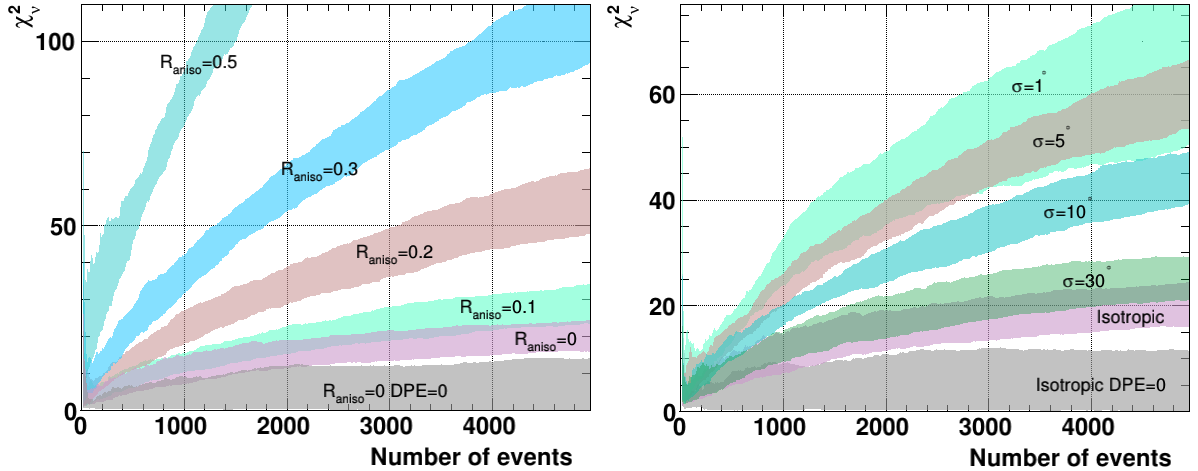}}
\caption{(Left) $\chi^2_\nu$ of anisotropic emissions with beam width of $5^\circ$ HWHM and various $ R_\textrm{aniso}$ (Right) $\chi^2_\nu$ of anisotropic emissions with $ R_\textrm{aniso}=0.2$ and various beam widths (HWHM). In both plots the grey band is for isotropic emission and no DPE and magenta band is for isotropic emission with DPE=20\%. 
\label{fig:sensitivity} }
\end{figure}

To assess the exposure required to identify an anisotropic  emission, data sets for different values of the anisotropy fraction ($R_\text{aniso}$) and of the beam HWHM ($\sigma$) are generated and tested against the null hypothesis. This is repeated 20 times for increasing number of events between $1-5000$.  The $\chi^2$ ranges obtained for the selected $R_{\textrm{aniso}}$ and $\sigma$ variations are shown in figure~\ref{fig:sensitivity}.

It is clear that $\lesssim 5 \times 10^3$ events are enough to distinguish anisotropic emission of as little as $R_\text{aniso}=0.2$ for beams with HWHM as large as $30^\circ$, over an isotropic emission pattern that contains the DPE contamination (see figure~\ref{fig:sensitivity} left). As expected, wider jets are harder to identify over the isotropic component, while below a width of $\sim 5^\circ$ the jet's inner structure is no longer resolvable (for $R_{\textrm{aniso}}=0.2$ and 60 initial photons, see figure~\ref{fig:sensitivity} right). 

The simplified statistical model presented here only provides a rough estimation of the expected sensitivity. We are now developing a more complete statistical model that will also include temporal anisotropy, treat the isotropic-DPE pattern as the null hypothesis, and address the challenge  of separating DPE events which have a jet-like structure.

We simulate two common used sources that emit isotropically, a 10 $\mu$Ci $^{137}$Cs, and a 2.7 $\mu$Ci $^{241}$AmBe, and apply an energy selection criterion which corresponds to the simulated number of photons/events $\pm20\%$ ($60\pm12$), this translates into an energy deposition of ($2.5 \pm 0.5$)\,keV (ER) and ($7 \pm 1.4$)\,keV (NR). The rate of events satisfying these selection criterion is found to be 625 events/day for ER and $1.25\times10^{4}$\ events/day for NR. Given these rates, a system that can operate stably for a few weeks could determine differences between ER events and NR events with the sort of anisotropies discussed above. When taking data of $^{241}$AmBe, the source will be shielded to reduce the level of gammas that will be introduced to the system.

% ========================================================================
\section{Experiment Operation}
% ========================================================================
\label{sec:exp_op}

\begin{figure}[htb]
\centerline{\includegraphics[width=0.85\linewidth]{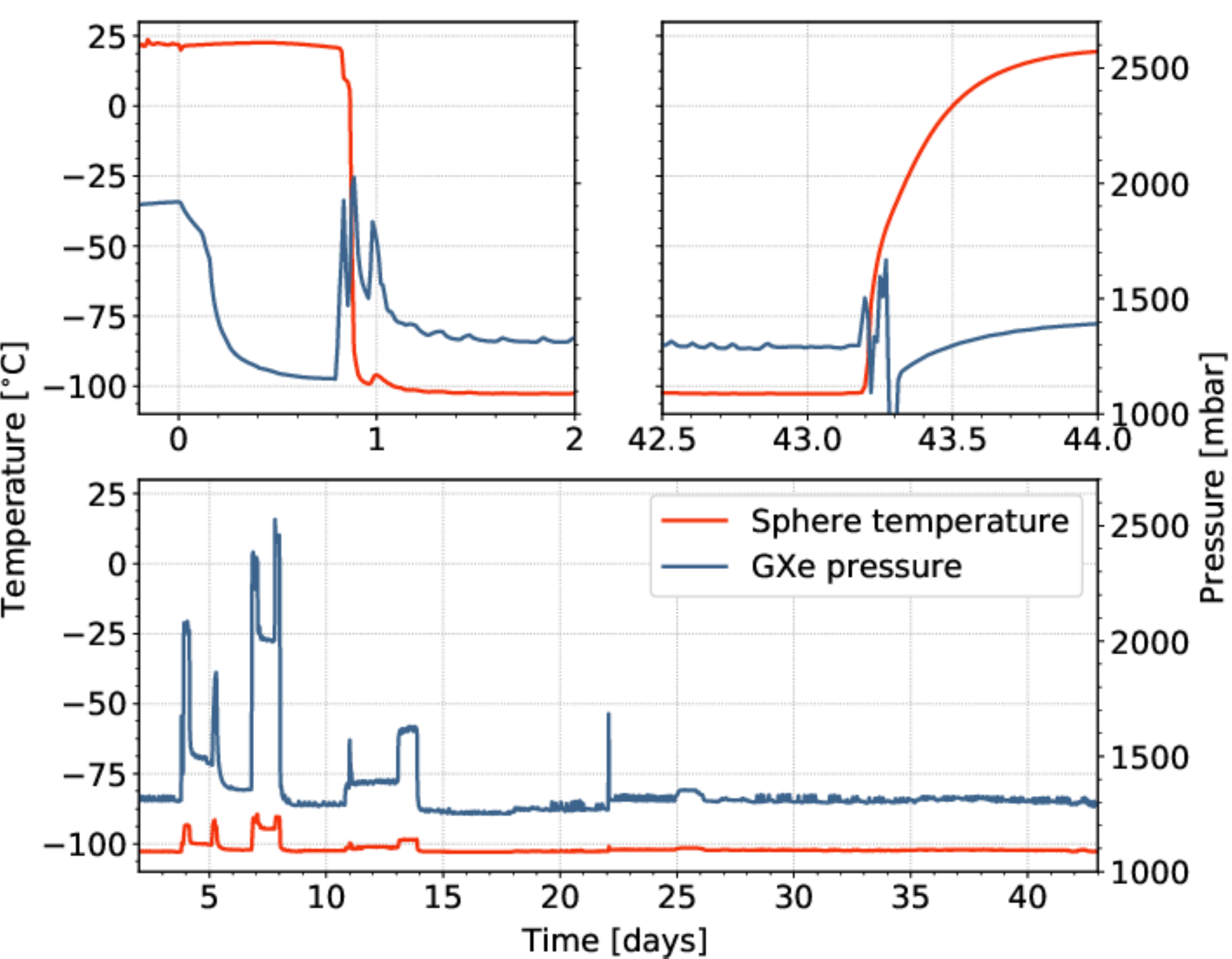}}
\caption{Sphere temperature and GXe pressure during the commissioning run. The cryo-cooler was turned on at Time = 0, when the system was already filled with $\sim 1900$ mbar GXe at room temperature. More GXe was added after about a day of initial cooling (top left). Stable and calm conditions in the sphere were reached after about 15 days (bottom). Large pressure spikes between days 2--14 are due to circulation pump turning on and off. Small hour-scale pressure spikes throughout the run are due to ambient air-conditioning controller algorithm, and they did not compromise the sphere's tranquility. After 43 days the cryo-cooler was turned off and the Xe inside was recuperated (top right).}
\label{fig:sc_CD18I}
\end{figure}

The system was tested  in a series of commissioning runs that included a full cycle of operation: $(a)$ system evacuation; $(b)$ xenon gas filling and liquefaction; $(c)$ xenon circulation and purification; $(d)$ detector operation; and $(e)$ xenon recovery. The slow-control readings for a typical operation cycle are shown in figure~\ref{fig:sc_CD18I}.

% ===========================
\subsection{Operation Cycle}
% ===========================
\label{subsec:op_cycle}

The IV is  evacuated  to reach a pressure of  $\sim5 \times 10^{-6}$ mbar and the OV is pumped to better than  $10^{-4}$ mbar, when cooling of the cold finger and filling of xenon is started. 
Cool-down of the IV takes a few hours, but it takes several days for the LXe filled detector to reach a thermodynamic steady-state since elements of the detector that are not in direct contact with LXe, such as the PMT assembly, cool mainly by radiation, which is considerably less efficient than conduction in this case. As described in section \ref{subsec:cryo_stability}, the system eventually reaches a state suitable for experimental-data acquisition. In the end of the run the xenon is cryo-pumped from the IV to the liquid-nitrogen-cooled reservoir tank.

% ===========================
\subsection{Cryogenic Stability}
% ===========================
\label{subsec:cryo_stability}
In order to have a reliable measurement of the temporal and spatial properties of the scintillation process, the LXe target region must be hydrodynamically and thermally stable. Phase transitions or turbulent flows in the xenon inside the sphere may create  inhomogeneities in the refractive index and damage correlation studies. Hence, the foremost milestone of \direxeno\ commissioning was to demonstrate that LXe can be stably maintained in the sphere as a single-phase fluid under laminar flow.

During the commissioning we have identified two dominant sources of thermodynamic disturbances that send GXe bubbles through the sphere when the system is filled with LXe. One is radiation from the OV that directly heats both the sphere and the LXe inside it, and the other is GXe bubbles that form inside the bottom reservoir and float up through the sphere to the LXe surface in the upper pools. Even though all the xenon containing parts of the detector are directly covered with ten-layered super-insulation (in addition to the super-insulation bag that covers the entire detector assembly), the sphere itself is exposed, and hence susceptible to radiative heating from the non-cold parts of the detector, such as the PMT assembly. This assembly is thermally disconnected from the IV and therefore cools very slowly. The steady-state temperature measured on the PMT assembly never goes below $-30^{\circ}$C. Nevertheless, the sphere itself, once cold, is massive enough to serve as a thermal buffer that prevents LXe from boiling inside it. Indeed, for a well isolated IV, this boiling inside the sphere stops several hours after its cool-down. The second source of bubbles that float through the sphere is eliminated by the special design of the bottom reservoir, which serves as a phase-separator and prevents bubbles from mixing into the up-flowing fluid when it is sufficiently cold (see section \ref{subsec:det_sys}). 

In a final commissioning run that extended over 44 days of continuous operation, we have shown that stable conditions with no visible bubbles inside the sphere can be reached within two weeks of cooling, and maintained for at least another 30 days. The SC parameters for this run are shown in figure \ref{fig:sc_CD18I} and demonstrate that the stability of the LXe inside the sphere is not compromised by ambient temperature fluctuations.
After reaching the 30-day stability milestone, the system was shut down for upgrades and preparation for science runs.

% ========================================================================
\section{First Measurements}
% ========================================================================
\label{sec:measur}

\begin{figure}[htb]
    \centerline{\includegraphics[width=1\linewidth]{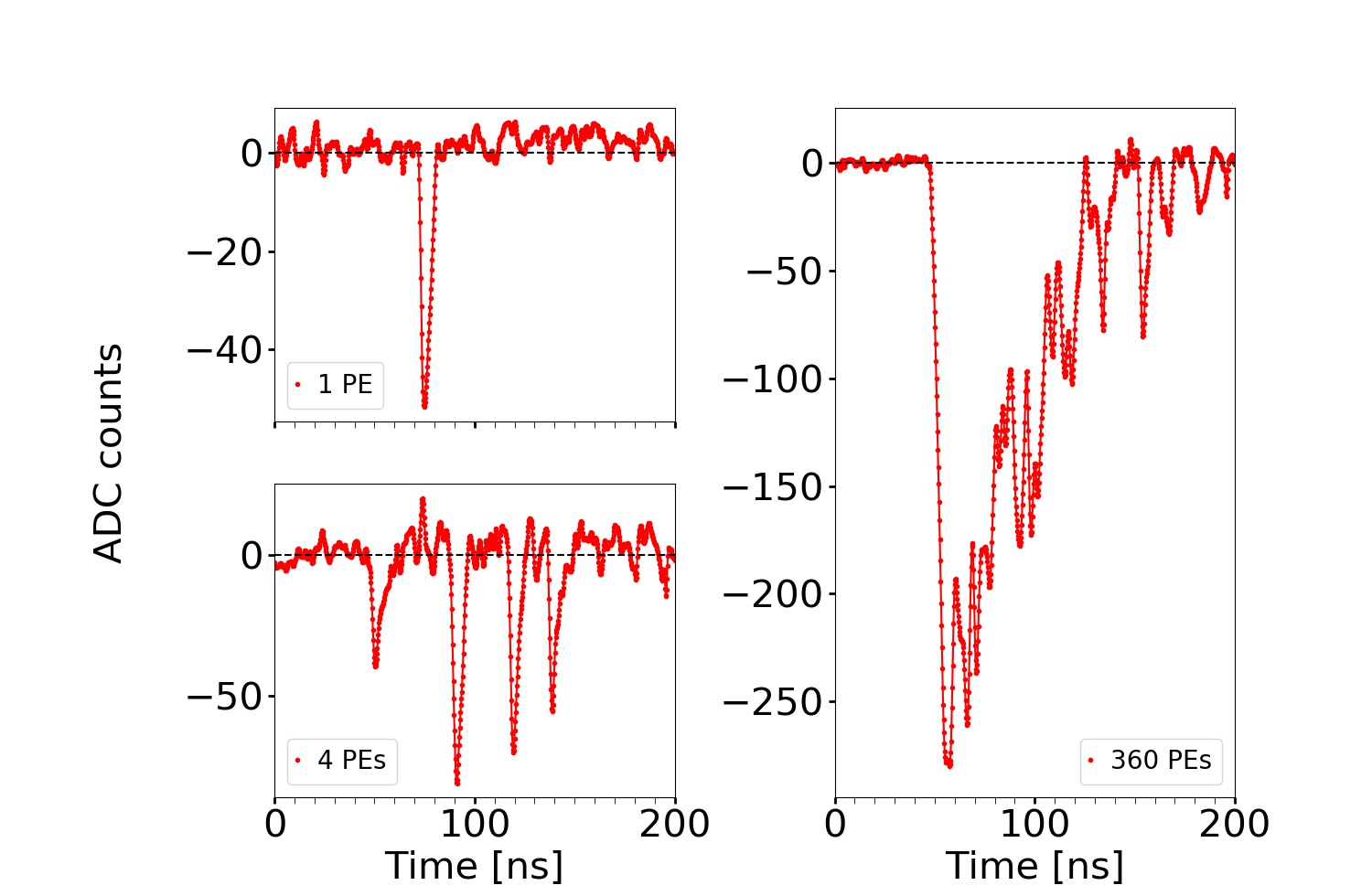}}
    \caption{Waveforms examples of one PE (top left), few PEs, (bottom left) and multiple PEs (right). }
    \label{fig:waveforms}
\end{figure}

The results reported here were collected in two commissioning runs. The setup  included the fully assembled detector with 19 PMTs, and the optical USB camera in place of the 20th PMT. Out of these, only 13 (17) PMTs were operational during the first (second) run due to connection and cabling issues inside the OV.  The trigger condition is a user-defined combination of PMT signal thresholds and their multiplicities. The events presented in this section were triggered by the crossing of a 100 mV threshold in at least three of the PMTs. In this trigger configuration the detector is not sensitive to low energy interactions, and for physics runs the trigger should be optimized. Some recorded waveforms are shown in figure~\ref{fig:waveforms}.

Data were collected  in several setups: $(a)$ background, without any external source; $(b)$ gain calibration with short UV-light pulses; $(c)$ $^{57}$Co $\gamma$ source exposure; and $(d)$ $^{137}$Cs $\gamma$ source exposure.

% ===========================
\subsection{Basic Analysis}
% ===========================
\label{subsec:basic}

The collected waveforms were analyzed offline to find their morphological and statistical features. These features were then used for event classification and rate calculations. Most notably, the \emph{baseline} and the \emph{peaks} were established. Since the anode rise time of the PMTs is 1.8\,ns, actual PEs can only create peaks wider than 3.6\,ns (FWHM). We therefore convolve the waveforms with a 2\,ns wide \emph{blackman filter}, to smooth-out narrow noise-related features before applying the peak detection algorithm.

% ===========================
\subsection{Gain Calibration}
% ===========================
\label{subsec:gain}

The PMT gains were measured in a designated gain calibration campaign which was performed before data acquisition of Background, $^{57}$Co, and $^{137}$Cs.

\begin{figure}[htb]
    \centerline{\includegraphics[width=1\linewidth]{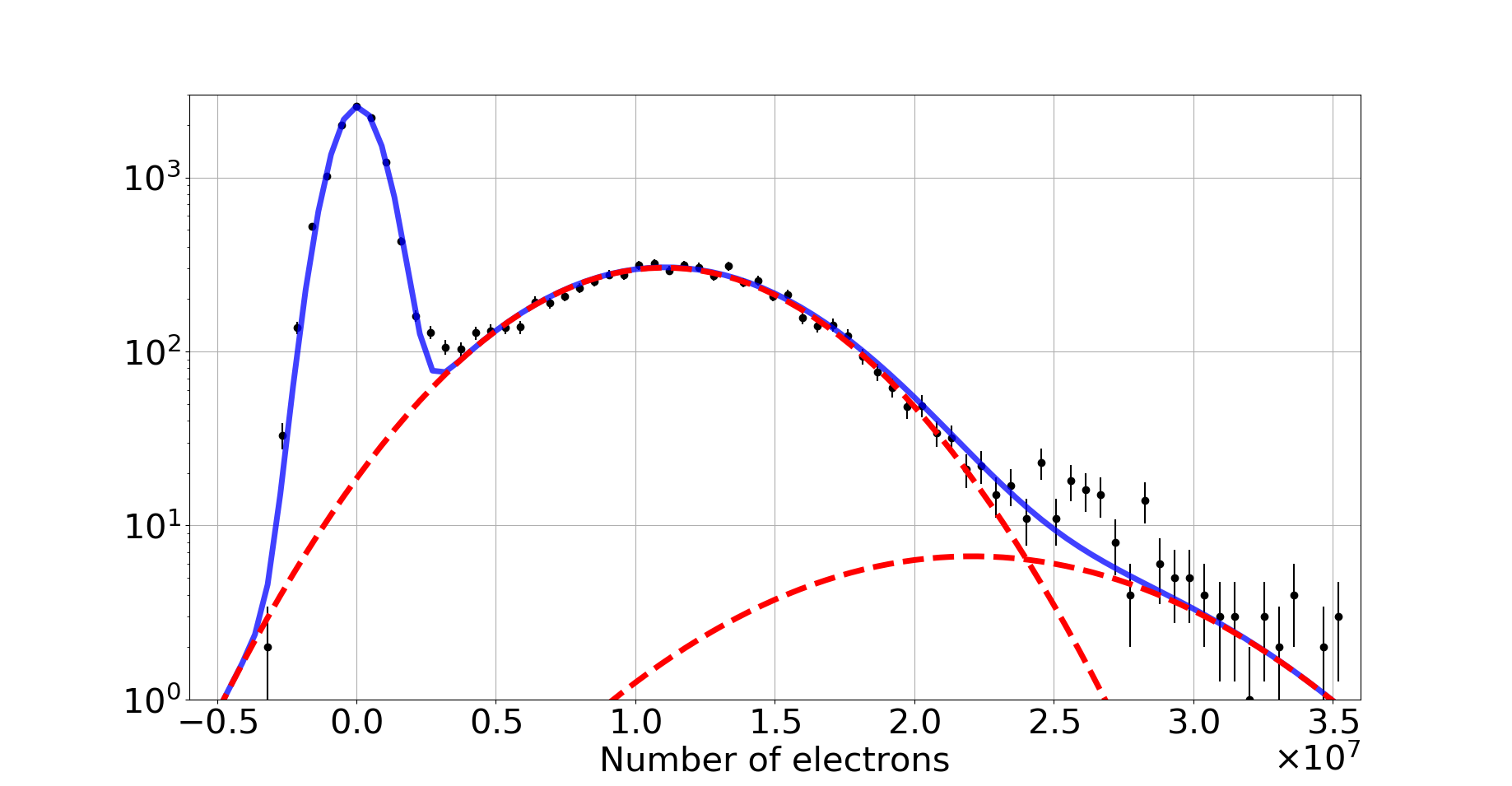}}
    \caption{Charge spectrum of a typical PMT at 850 V under pulsed-LED illumination (black crosses) and a combined fit (blue) of the pedestal, one PE and two PE components. Gain is $\sim 11 \times 10^6$.}
    \label{fig:gain}
\end{figure}

The UV pulsed-LED described in section~\ref{subsec:det_sys} was set to generate a fast ($\lesssim 1$\,ns) and faint calibration signal (365 nm) for the PMTs. For each PMT a sample of $10^5$ such events was used to produce the one PE area spectrum, which was then fitted by an empirical distribution model. Assuming normal underlying PE distributions, the combined model can be written as

\begin{equation}
\begin{split}
    g(x) = A_0 G\left( x; \mu_0, \sigma_0 \right) +& A_1 G\left(x; \mu_0 + \mu_1,\sqrt{\sigma_0^2 + \sigma_1^2}\right) \\ +& A_2 G\left(x; \mu_0 + 2\mu_1,\sqrt{\sigma_0^2 + 2\sigma_1^2}\right)
\end{split},
\end{equation}
where $G(x; \mu, \sigma) = \exp(-(x-\mu)^2\, /\, 2\sigma^2)$ is a Gaussian shape function. The first term represents the pedestal, which is mainly due to electronic noise. The second and third terms are the one and two PE charge distributions, respectively. The effects of higher order terms as well as DPE (at $\lambda =365$\,nm) are negligible with respect to the one PE and therefore not considered. An example of a typical spectrum with its fit is given in figure~\ref{fig:gain}. The charge resolution ($\sigma_1$/$\mu_1$) of the active PMTs is $0.48$ with a spread of (std) $ 0.08$.

% ===========================
\subsection{Temporal Calibration \& Resolution}
% ===========================
\label{subsec:temp_res}

The temporal response of each data channel in the detector, from incident photons at the PMT window to the digitizer output, depends on the internal response of the PMT itself and on the electronic noise and latency fluctuation introduced by the DAQ components. Calibration of the relative delay between the channels is therefore necessary for a correct temporal analysis of events.

In each of the PMTs, when the photocathode is illuminated with photons over its entire area, the transit time of a PE pulse inside the PMT has a fluctuation, a \emph{transit-time spread} or jitter, which is inversely proportional both to the square root of the PE number, and to the square root of the operation voltage \cite{pmthandbook2007}. The manufacturer reports an intrinsic jitter of 0.8\,ns (FWHM) at 800\,V for our PMTs, to which the channel electronic fluctuation should be added. 

When two PMTs record photons that were emitted simultaneously from the center of the detector (i.e., at equal photon time-of-flight), the time delay between the two sets of readings at the DAQ is expected to be normally distributed, and the width of the distribution is the detector's (pairwise) uncertainty in timing. This is also known as the \emph{coincidence time resolution} (CTR), and is given by
\begin{equation}
	\tau_{12}^2 = \tau_{1}^2 + \tau_{2}^2,
\end{equation}
where $\tau_{1}$ and $\tau_{2}$ are the overall jitters of the two PMTs. Due to the intrinsic jitter of the PMTs, we expect $\tau_{12} \gtrsim 1.13$ ns (FWHM).

\begin{figure}[htb]
    \centerline{\includegraphics[width=1\linewidth]{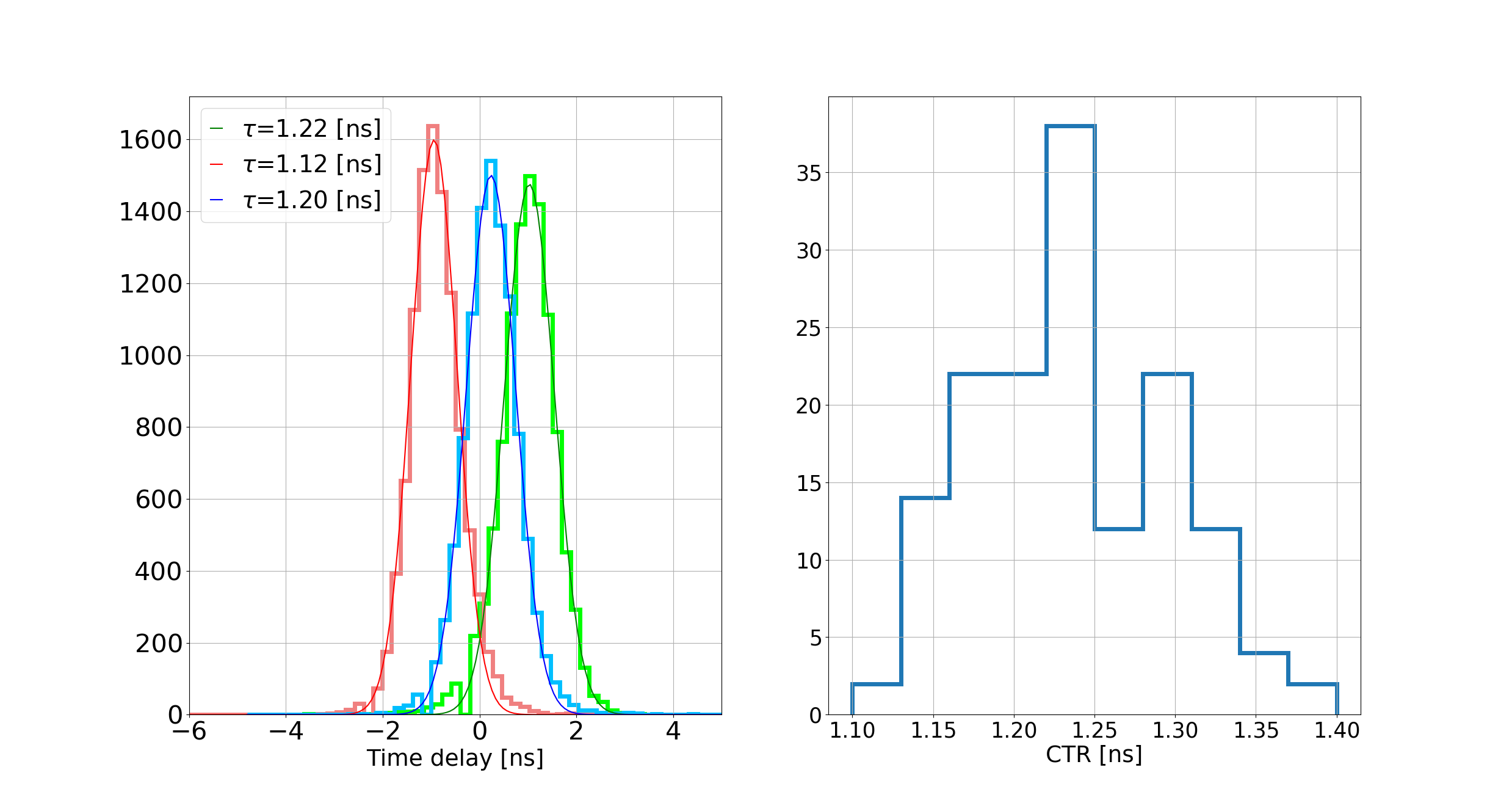}}
    \caption{The delay distribution between three pairs of PMTs with a Gaussian fit; $\tau$ is the FWHM taken from the fit (left). The distribution of the CTR ($\tau$) from all pairs of PMTs (right).}
    \label{fig:delay}
\end{figure}

We used high occupancy ($> 2000$ PEs per waveform) ER events generated by $^{137}$Cs $\gamma$ radiation to estimate the CTR of the detector. These events can be considered as point sources with nearly spherically symmetric emission. The first light of each waveform in each event was timed at the point it reached $10\%$ of the pulse maximum amplitude, relative to the baseline, and the times were used to find the pairwise-delay distributions of the PMTs. The delay distributions of three typical PMT pairs and the distribution of the CTR between all pairs of PMTs are shown in figure~\ref{fig:delay}. The time uncertainty of the detector is taken to be the largest CTR (i.e., the worse resolution). In these conditions it was found to be $\lesssim 1.4$\,ns (FWHM), which is sufficient for measuring the temporal structure of LXe scintillation and detecting superradiance (possibly a sub-ns process).

% ===========================
\subsection{Scintillation Spectrum \& Light Yield}
\label{subsec:sprc&Ly}
% ===========================

\begin{figure}[htb]
\centerline{\includegraphics[width=1\linewidth]{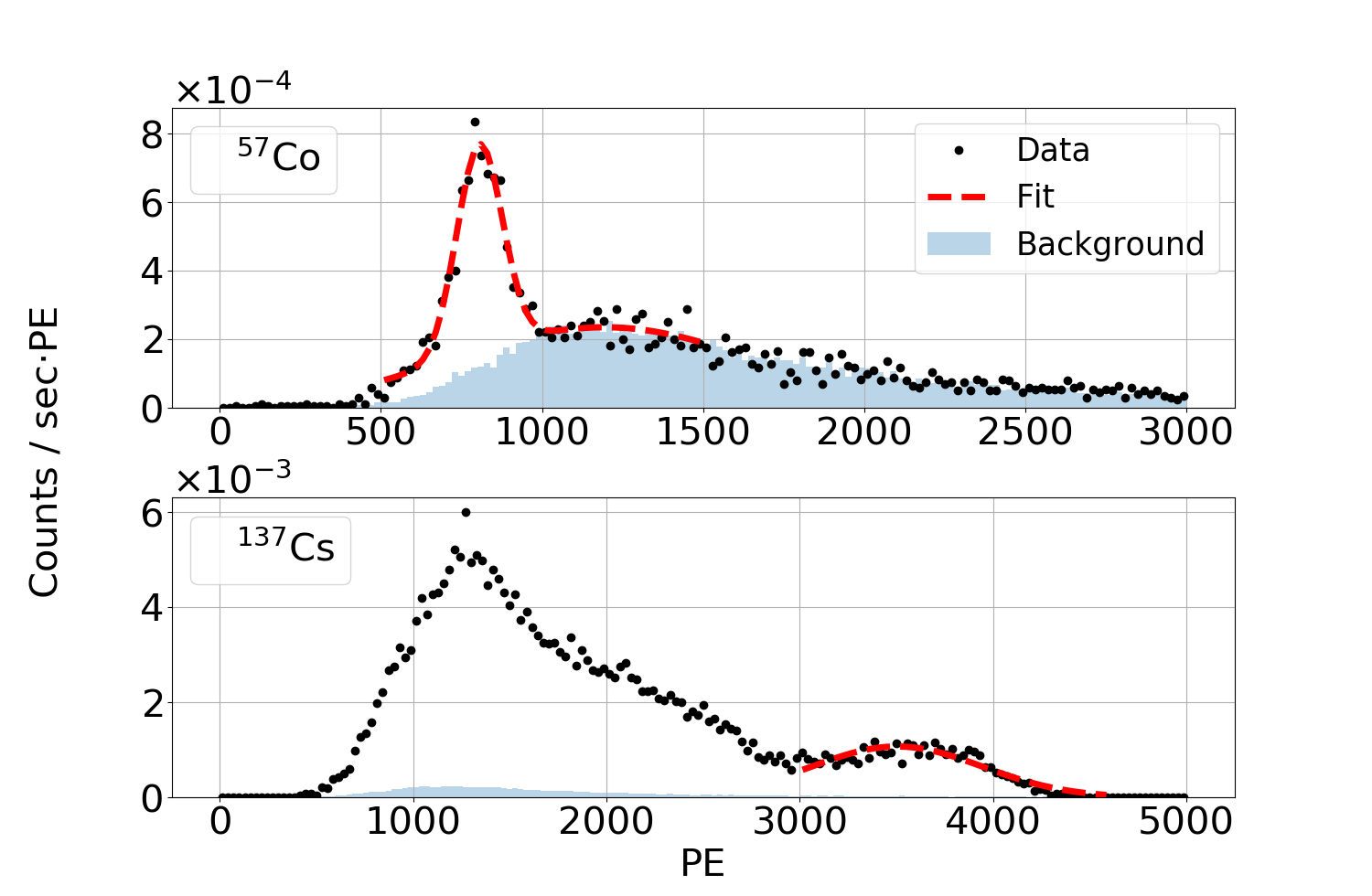}}
\caption{(top) Scintillation spectrum of $^{57}$Co and background. The full absorption peak is measured to be ($806 \pm 71)$\,PE. (bottom) The spectrum of $^{137}$Cs, the peak seen at $\sim1200$PE is the Compton shoulder whereas the one at ($3503 \pm 441$)\,PE is the full absorption. The background data set is identical in both cases and is taken from the ``background only'' run.}
\label{fig:Co57Cs137}
\end{figure}

The $^{57}$Co scintillation light spectrum is shown in figure~\ref{fig:Co57Cs137} (top). A Gaussian fit to the dominant peak of 122\,keV $\gamma$-rays gives ($805 \pm 71$)\,PE. These results correspond to a light yield of $L_\text{y} = (6.6 \pm 0.6)$~PE~/~keV and an energy resolution of 9\%. Hence, for the full experiment setup with 20 PMTs (which was not tested in this run), a light yield of about $L_\text{y} \sim (7 \pm 1)$~PE~/~keV is expected. The \emph{light collection efficiency} (LCE), defined as the ratio between the number of photons reaching the PMT photocathode to the number of photons emitted, is given by
\begin{equation}
    L_\text{c} = \frac{N_\textrm{PE}(\epsilon)}{N_\gamma(\epsilon) \cdot PMT_\textrm{eff} \cdot \textrm{P}_{\textrm{DPE}} }
    , 
\end{equation}
where $N_\gamma$ is the absolute scintillation light yield for incident particles of energy $\epsilon$, $N_\text{PE}$ is the number of PEs measured by the PMTs per scintillation event, $PMT_\text{eff}$ is the PMT total efficiency which accounts both for the QE and CE, and $\textrm{P}_{\textrm{DPE}}$ is the probability for DPE production (20\%). For QE = 0.33, CE=0.81(see section ~\ref{sec:det_sens}), and $N_\gamma = 65 \pm 2$~Photons~/~keV \cite{NEST,Szydagis:2011tk} (with zero electric field), the LCE of the tested detector configuration is $L_\text{c} = (0.31 \pm 0.04)$. For comparison, the geometrical coverage of 17 PMTs gives $\sim\!0.35$. The difference between these two values can be attributed to waveform tails not included in the digitization window and to the absorption of photons in the sphere ($\sim2\%$) and LXe ($\sim0.5\%$).  

Table~\ref{tab:light_yield} shows a summary of the light yield and energy resolution for $^{57}$Co and $^{137}$Cs sources (see spectrum in figure~\ref{fig:Co57Cs137} bottom).

\begin{table}[htb]
    \centering
    \caption{Light yield and energy resolution for different $\gamma$-ray sources with 17 PMTs.}
    \begin{tabular}{|c||c|c|c|c|}
        \hline
        Source & Energy [keV] & Light yield  [PE / keV]  & \thead{Energy resolution\\ ($\sigma_\text{E}$ / E) [\%]} \\
        \hline
        \hline
        $^{57}$Co & 122 & 6.6 $\pm$ 0.6 & 9 \\
        $^{137}$Cs & 662 & 5.3 $\pm$ 0.7 & 13 \\
        \hline
    \end{tabular}
    \label{tab:light_yield}
\end{table}

% ========================================================================
% summary
% ========================================================================
\section{Summary}
\label{sec:sum}

We have constructed the setup of \direxeno, an apparatus designed to measure the spatial and temporal distributions of LXe scintillation and demonstrated the sensitivity of the detector to non isotropic emission in the form of a beam with as little as $ \textnormal{R}_\textrm{aniso} \gtrsim 20\% $ and HWHM as large as $\theta \lesssim 30^\circ$. A run--time of several weeks is required if using typical radioactive sources. The detector has shown stable conditions for a run--time of 44 days. The time resolution on calibration events was measured to be $\lesssim 1.4$\,ns (implying $\lesssim 0.55$\,ns 1$\sigma$) and the energy resolution at 122\,keV was measured to be 9\%. These values are sufficient to measure effects like super-radiance or other non-linear scintillation processes. 

Once we will improve the data quality from the detector by utilizing all 20 PMTs and better understanding the detector response, we will collect data from the same $\gamma$ sources in addition to a neutron source.

The impact of the discovery of an anisotropy or other new behavior in the emission of scintillation on, for example, dark matter searches, will depend on the nature and observability of the new phenomenon. Searches for effects of this kind have not been reported by previous experiments. The performance of the current setup is designed to give optimal sensitivity to correlation in time and space among scintillation photons.

\section{Acknowledgment}
\label{sec:ack}
The authors thank A. Breskin, V. Chepel, L. Arazi and D. Vartsky for useful discussions, and  S. Shchemelinin for assistance with transparency measurements.
The authors thank the Weizmann Institute for the generous support. This work was supported by a Pazy-Vatat grant for young scientists and by the I-CORE Program of
the Planning Budgeting Committee and the Israel Science Foundation (grant No. 1937/12). RB is the incumbent of the Arye and Ido Dissentshik Career Development Chair.

% ========================================================================
% bibliography
% ========================================================================
\bibliographystyle{JHEP}
\bibliography{Direxeno_v2_Bib}
\end{document}